\documentclass[12pt,preprint]{aastex}

\shorttitle{Close binary stars.~XII}
\shortauthors{Pribulla \& et al.}

\begin{document}

\title{Radial Velocity Studies of Close Binary
Stars.~XII\footnote{Based on the data obtained at the David Dunlap
Observatory, University of Toronto.}}

\author{Theodor Pribulla}
\affil{Astronomical Institute, Slovak Academy of Sciences \\
059 60 Tatransk\'a Lomnica, Slovakia}
\email{pribulla@ta3.sk}
%\and
\author{Slavek M. Rucinski, George Conidis,
        Heide DeBond, J.R. Thomson}
\affil{David Dunlap Observatory, University of Toronto \\
P.O.~Box 360, Richmond Hill, Ontario, Canada L4C~4Y6}
\email{(rucinski,conidis,debond,jthomson)@astro.utoronto.ca}
%\and
\author{Kosmas Gazeas}
\affil{Department of Astrophysics, Astronomy, and Mechanics\\ National and 
Kapodistrian University of Athens, 157 84 Zographou, Athens, Greece}
\email{kgaze@skiathos.physics.auth.gr}
%\and
\author{Waldemar Og{\l}oza}
\affil{Mt. Suhora Observatory of the Pedagogical University\\
ul.~Podchora\.{z}ych 2, 30--084 Cracow, Poland}
\email{ogloza@ap.krakow.pl}

\begin{abstract}
Radial-velocity measurements and sine-curve fits to the orbital radial
velocity variations are presented for ten close binary systems: OO~Aql,
CC~Com, V345~Gem, XY~Leo, AM~Leo, V1010~Oph, V2612~Oph, XX~Sex, W~UMa,
and XY~UMa. Most of these binaries have been observed spectroscopically 
before, but our data are of higher quality and consistency 
than in the previous studies.

While most of the studied eclipsing pairs are contact binaries,
V1010~Oph is probably a detached or semi-detached double-lined binary
and XY~UMa is a detached, chromospherically active system whose
broadening functions clearly show 
well defined and localized dark spots on the primary component. 
A particularly interesting case is XY~Leo, which is
a member of visually unresolved quadruple system composed of a contact
binary and a detached, non-eclipsing, 
active binary with 0.805 days orbital period. 
V345~Gem and AM~Leo are known members of visual
binaries. We found faint visual companions at about 
2--3 arcsec from XX~Sex and XY~UMa.
\end{abstract}


\section{INTRODUCTION}
\label{sec1}

This paper is a continuation of a series of papers (Papers
I -- VI and VIII -- XI) of radial-velocity studies of close binary stars
and presents data for the eleventh group of ten close binary stars observed
at the David Dunlap Observatory. For full references to the
previous papers, see \citet[ Paper XI]{ddo11}; 
for technical details and conventions, presentation of the
broadening functions approach and for preliminary 
estimates of uncertainties, see the interim summary 
paper \citet[ Paper VII]{ddo7}. The recent DDO studies
use the very efficient program of \citet{pych2004} 
for removal of cosmic rays from 2-D images.

While most of the data used in this paper
were determined using the broadening functions 
(from now on called BF's) extracted -- as in the previous
papers -- from the region of the Mg~I triplet at 
5184~\AA, we also used a few observations of
XY~UMa from a region centered at 6290~\AA. This 
experimental setup, which included telluric features, 
was used (i)~because of concerns about flexure effects 
in our spectrograph and (ii)~to improve visibility of the late-type
secondary component in this binary.
The experiment provided a good check on the stability of our radial-velocity 
system and -- to a large extent -- alleviated our concerns. 
We found also that the stellar lines around 6290~\AA\ 
and 6400~\AA\ were generally too weak to replace the 5184~\AA\ feature 
for routine stellar BF determinations. The BF's for XY~UMa extracted from the 
6290~\AA\ were more noisy than those from the 5184~\AA\ spectral region
and the detection of the secondary component was not improved.

In August 2005, a new grating with 2160 l/mm was acquired to replace
the previously most frequently used 1800 l/mm grating which after
many years of use lost its efficiency. This markedly improved
quality of the observed spectra and of the resulting BF's. 
The older grating was used only for 2005 observations of V1010~Oph and XY~UMa.

The radial velocity (hereafter RV) observations reported in this paper
have been collected between April 2005 and April 2006. The ranges of dates 
for individual systems can be found in Table~\ref{tab1}.
Selection of the targets in our program remains quasi-random: At a given time,
we observe a few dozen close binary systems with periods usually
shorter than one day, brighter than 10 -- 11 magnitude and with
declinations $>-20^\circ$; we publish the results in groups of ten systems
as soon as reasonable orbital elements are obtained from measurements evenly
distributed in orbital phases. In this paper we re-observed several relatively
bright systems (V1010~Oph, W~UMa, XY~Leo) to ascertain possible systemic
velocity changes which could indicate presence of a third body in the system.
Similarly as in our previous papers dealing with 
spectroscopically multiple systems 
(here the cases of XY~Leo and V345~Gem), RV's for the eclipsing pair 
were obtained from BF's with the third-star sharp peaks removed first, 
as described most recently in \citet{ddo11}.

As in other papers of this series,
whenever possible, we estimate spectral types of the program stars
using our classification spectra. These are compared with the mean $(B-V)$ 
color indices usually taken from the Tycho-2 catalog \citep{Tycho2} and 
the photometric estimates of the spectral types using the relations 
published by \citet{Bessell1979}.

This paper is structured in a way similar to that of previous papers, in that 
most of the data for the observed binaries are in two tables consisting of the 
RV measurements in Table~\ref{tab1} and the sine-curve orbital solutions
in Table~\ref{tab2}. The
RV's and the corresponding spectroscopic orbits for all ten systems 
are shown in phase diagrams in Figures~\ref{fig1} to \ref{fig3}. 
The RV's are fitted without proximity
effects taken into acoount. This results in systematic deviations of the
fits close to the eclipses. A further improvement of the orbits
can be obtained by simultaneous fitting of the RV's and photometry taking into
account the proximity effects end eclipses, but it has not been
attempted in this paper. 
The measured RV's are listed in Table~\ref{tab1} together with weights, 
determined from $1/\sigma^2$, as based on individual determinations of centroid
velocities. This weighting scheme, which accounts for differences
in the {\it relative\/} quality of observations, markedly improves the 
overall quality of the orbital solutions. However, these errors -- resulting 
from non-linear least-squares fitting -- tend to stay at a level
of a few 0.1 km~s$^{-1}$ and therefore under-estimate the
real uncertainties. In turn, the errors of the unit weight, as given 
by the fit (Table~\ref{tab2}, column $\epsilon_i$), 
combine the errors of the individual RV's with 
all systematic deviations (proximity effects, flexures of the spectrograph,
mismatch of template spectral types, etc.) and thus over-estimate
the measurement uncertainties.

Table~\ref{tab2} contains also 
our new spectral classifications of the program
objects. Section~\ref{sec2} of the paper contains brief summaries of
previous studies for individual systems and comments on the new data.
Examples of BF's of individual systems extracted from spectra observed
close to quadratures are shown in Fig.~\ref{fig4}.

The data in Table~\ref{tab2} are organized in the same manner as in 
the previous papers of this series. 
In addition to the parameters of spectroscopic orbits, 
the table provides information about the relation between the 
spectroscopically observed upper conjunction of the more massive 
component $T_0$ (not necessarily identified with the 
primary, i.e.\ deeper eclipse) 
and the recent photometric determinations of the primary
minimum in the form of the $O-C$ deviations for the number of 
elapsed periods $E$. For XX~Sex, the reference ephemeris was 
taken from \citet{wils2003}; for the rest
of the systems, the ephemeris given in the on-line 
version of ``An Atlas O-C diagrams of eclipsing binary 
stars''\footnote{http://www.as.wsp.krakow.pl/ephem/} 
\citep{Kreiner2004} were adopted. Because the on-line 
time-of-eclipse data are 
frequently updated, we give those used for the computation of 
the $O-C$ residuals below Table~\ref{tab2} (status as of May 2006). 
The deeper eclipse in W-type contact binary systems corresponds to 
the lower conjunction of the more massive 
component; in such cases the epoch in Table~\ref{tab2} is a 
half-integer number.

\section{RESULTS FOR INDIVIDUAL SYSTEMS}
\label{sec2}

\subsection{OO~Aql}
\label{ooaql}

This bright ($V_{max} = 9.50$) contact binary 
is quite unusual in that it has
a mass ratio close to unity in spite of being an A-type 
system. It also shows a discrepancy between the
spectral type and the color index. While \citet{roma1956} assigned the
G5V spectral type to the system, 
\cite{hill1975} found the K0 type based on the classification spectra. 
The observed color indices $(B-V) = 0.76$ \citep{egge1967} and 
$(b-y) = 0.46$ \citep{ruka1981} indicate a late spectral type of
G8 to K0, but as pointed by \citet{egge1967}, the 
reddening in this galactic direction is very patchy and may
reach $E_{B-V} \simeq 0.15$. 
Our classification spectra give discordant estimates: 
While the G-band (4300~\AA) gives about F9V, 
the hydrogen lines are weak indicating a late G type
perhaps G8V.

In spite of its relatively high brightness, 
the system was not observed by the Hipparcos satellite so no
direct measure of the distance is available. Using \citet{rd1997}
calibration and assuming a wide range of spectral
types of F8V to G8V, we obtain a range of absolute magnitudes
of  $M_V = 3.06$ (F8V) to $M_V = 3.66$
(G8V), corresponding to minimum distances (no reddening) of
194 and 147 pc, respectively. The reddening of 
$E_{B-V} = 0.15$ would increase these estimates to 
208 and 158 pc, respectively. 

\citet{moch81} suggested that OO~Aql may be considered as
a prototype of a sub-group of contact binaries with
components recently evolved into contact after 
a considerable angular momentum loss 
in the pre-contact stage. The view that the system represents a rare, 
transitional phase in the evolution of contact binaries
was later shared by \citet{hriv1989} who
presented a consistent, combined radial-velocity
and light-curve solution, and showed 
that the orbital inclination of the system is close to 90\degr.
A sine-curve approximation to the radial velocities 
obtained with the cross-correlation method led to a 
spectroscopic orbit with $V_0  = -46.4 \pm 0.9$ km~s$^{-1}$, 
$K_1 = 147.3 \pm 1.4$ km~s$^{-1}$, and 
$K_2 = 178.5 \pm 112.0$ km~s$^{-1}$. This resulted
in a large mass ratio of $q = 0.825 \pm 0.012$, which 
-- with the proximity effects included -- raised  
to $q = 0.843 \pm 0.008$. This value is very close
the one obtained from our new spectroscopic 
orbit, $q = 0.846 \pm 0.007$. 

The center-of-mass velocities
of \citet{hriv1989} and the present result differ by about 7 km~s$^{-1}$.
In the view of typical differences found for contact binaries
from analyses of different authors
\citep{prib2006}, we regard this as manifestation of a systematic effect 
which could be caused by differences in radial-velocity standard
systems or/and differences in methods used for radial-velocities
determination (cross-correlation or broadening functions, combined
with the RV determination via centroids, Gaussians or rotational profiles).

We see the OO~Aql system practically edge-on, so the true masses are 
very close to the projected ones. 
With $(M_1+M_2) \sin^3 i = 1.954 \pm 0.019 M_\odot$
and the new mass ratio, we obtain $M_1 = 1.058 \pm 0.011 M_\odot$ and
$M_2 = 0.895 \pm 0.009 M_\odot$. The mass of the primary component
corresponds to the main sequence spectral type G1V 
(the secondary component would be G6V, if not in contact). Thus,
the primary spectral type, as estimated from its mass, is close
to the hot end of the two extremes in the direct estimates, 
F8V and G8V. The distinctly red color of the system, 
$(B-V) = 0.76$, would then imply a surprisingly large reddening of 
$E_{B-V} \simeq 0.2$.

\subsection{CC~Com}
\label{cccom}

CC Com is a totally eclipsing contact binary at the extreme short-period 
end of the currently available period distribution.
With its period of only 0.22068 days, it has been a record holder 
for a long time until a contact binary with 
the 0.215 day period was found in 
47~Tuc by \citet{weld2004}. Because of its extreme
properties, it has been a subject of several  
photometric studies (e.g., \citet{rucin1976,linnell89}) and of two 
spectroscopic studies \citep{rucin1977,lean1983}.

Using an old, now totally obsolete 
technique of measuring individual metallic 
lines in image-tube, 4m-telescope spectra, \citet{rucin1977} determined  
relatively reasonable spectroscopic parameters: 
$V_0 = -10.2 \pm 5.4$ km~s$^{-1}$, $K_1  = 122.0 \pm 5.5$
km~s$^{-1}$, $K_2 = 235.9 \pm 4.8$ km~s$^{-1}$ confirming 
the photometric mass ratio found from
the timing of the eclipse inner contacts \citep{rucin1976}. 
The broadening functions do not show any trace of a third component, 
hence we regard a 7.3 km~s$^{-1}$ difference between systemic velocities
of \citet{rucin1977} and present paper as resulting from systematic
errors. A spectroscopic orbit of \citet{lean1983}, based
on a few rather poor measurements, while agreeing with the 
above, has been of a limited use.

A new determination of radial velocities was a real challenge 
for our 2m-class telescope due to the relatively 
low apparent brightness ($V_{max} = 11.0$), red color
and the very short period of the system. 
Even with relatively long exposures of 500s (2.6\% of the orbital
period) the spectra were very noisy. Moreover, with the K5V 
spectral type, the system is relatively faint at the Mg~I triplet. 
We solved these difficulties by re-observing the quadrature
segments of the orbit several times with the total number of 
134 spectra. As expected, individual BF's were
poor so we sorted them in the phase domain and then used
temporal smoothing with a $\sigma = 0.02$ Gaussian filter 
with the subsequent re-binning to a step of 0.02 in phase. 
Consequently, Table~\ref{tab1} 
gives radial velocities in equidistant phases with the mean values
of the HJD time equal to the average time 
of the contributing observations. Our new spectroscopic 
elements are within the errors of those determined by \citet{rucin1977}. 
Total minimum mass of the system, 
$(M_1+M_2) \sin^3 i = 1.088 \pm 0.014 M_\odot$, is expected to be
close to the true value because the orbit of the system is 
seen practically edge-on.

\subsection{V345~Gem}
\label{v345gem}

The photometric variability of V345~Gem was discovered 
by the Hipparcos satellite \citep{hip}, where it was 
catalogued as a periodic variable with a period of 
0.1373890 days. Later, \citet{duer1997} classified 
the system as a pulsating 
variable on the basis of the period-color relation. 
V345~Gem was subsequently included in the GCVS \citep{namelist74} 
as a $\delta$~Scuti variable. Finally, a high-precision 
photometry of \citet{gomez2003} (with
both components of the visual pair within the photometric
aperture -- see below) showed that system is 
very probably a contact binary with twice the period (0.274778 days)
and a photometric variation amplitude of 0.07 mag. 
These authors determined the first 
reliable ephemeris $Min.I = BJD 2448362.7224(10) + 
0.2747736(2)\times E$ by doubling of the Hipparcos period.

V345~Gem is the member of visual binary WDS~07385+3343 \citep{wds} 
consisting of components with magnitudes
$V_1= 8.08$ and $V_2 = 9.35$, separated 
by 3.1 arcsec and at present positioned 
practically perpendicular to our spectrograph slit. 
\citet{namelist74} commented 
that the photometric variability of V345~Gem might be due to 
the fainter component. In fact, it was the early spectral type of 
the dominant component (F0) which resulted in an incorrect
classification of the star as a pulsating variable 
by \citet{duer1997}.

By mistake, our spectroscopic observations of V345~Gem were first 
focused on the primary component of the visual pair. After some 
time it became obvious that the primary component is a
single, slowly-rotating star and that the contact binary has to
be identified with the fainter companion. The presence of the
bright companion remained still obvious in the spectra of the
close binary because -- due to the relatively 
poor seeing at the DDO site of typically 1--4 arcsec -- 
the spectra of the fainter component were always contaminated by the 
visual companion. Although the companion spectral signature could be
removed by fitting three Gaussian profiles to the extracted BF's, some 
persistent features most likely caused by the different level 
and slope of the continua in both stars did remain. 
In spite of these difficulties, the
resulting spectroscopic orbit of the contact pair is of a good 
quality with the minimum mass of 
$(M_1+M_2) \sin^3 i = 1.054 \pm 0.013 M_\odot$, which is rather 
high for its orbital period of 0.275 days 
and the mass ratio of $q = 0.143$. 
After a correction for the third light of the brighter 
visual companion, the photometric amplitude of the 
contact pair is about 0.33 mag, which implies a 
high inclination angle so that the total mass is
probably close to the above minimum-mass estimate.

The recent secondary minimum ($HJD 2,453,731.9423$, \citet{nelson2006}) 
coincides in phase with the upper spectroscopic conjunction of 
the more massive component. Hence, V345 Gem is a contact binary of 
the W sub-type.

The Hipparcos trigonometric parallax $\pi = 8.61 \pm 1.77$ mas may 
be affected by the visual binary character of
the wide pair. If we take F7V spectral type, estimated 
for the fainter component of the visual pair from our 
classification spectra and the corresponding $(B-V)_0 = 0.50$, 
the absolute magnitude calibration of \citet{rd1997} 
gives $M_V = 4.12$ which is very close to the absolute magnitude
$M_V = 4.03$ determined from the parallax and the visual magnitude.

The radial velocity of the visual companion, 
$V_3 = -1.68 \pm 0.15$ km~s$^{-1}$, 
determined from strongly contaminated spectra of the binary,
$L_3/(L_1+L_2)>0.10$, is
close to the center of mass velocity of the contact binary,
$V_0 = 0.03\pm 0.68$ km~s$^{-1}$; however, this value may be affected
by the asymmetric distribution of the third light across the 
spectrograph slit. A similar value, 
$V_3 = -2.55 \pm 0.73$ km~s$^{-1}$, was found from the 
44 spectra of the third star which were observed by mistake,
but were well centered on the spectrograph slit. 
No RV variations of the brighter visual component have 
been found. This confirms the physical bond of the visual pair and 
indicates that the orbital motion in the visual orbit is very slow.
Radial velocities of the bright visual component of V345~Gem 
are available in \ref{tab3}.

\subsection{XY~Leo}
\label{xyleo}

The bright contact binary XY~Leo is a member of a quadruple system 
with an active binary and a contact binary
on 20-year period mutual orbit \citep{bard1987}. The system
has a long history of being recognized as somewhat unusual
and abnormally bright in the X-rays \citep{crudd1984}
and the chromospheric Mg~II emission \citep{rucin1985}.

The multiple nature of the system had been first indicated by the
periodic changes of the orbital period of XY~Leo; the
light-time effect interpretation and the expected
nature of the third body was extensively discussed by 
\citet{gehl1972}. The authors deduced the minimum mass
for the third body to be about 1~$M_\odot$. 
\citet{struve1959} obtained 
the first spectroscopic orbit for the contact binary and noted 
strong Ca~II H and K emissions, which were considered to originate
on the more massive component. Finally, \citet{bard1987} found 
the companion spectroscopically as a BY~Dra-type binary of a mid-M
spectral type with its own short orbital period of 0.805 days. 
The lines of this component were seen as narrow absorption 
features in the red spectra, 
relatively easy to measure compared with the spectral
lines of the contact binary. The light contributions of the third 
and fourth components at $H_\alpha$ were found to be 
7.5\% and 2.5\% of the total light, respectively. 
The orbits of both systems are not coplanar: 
While the inclination 
of the eclipsing pair is about 66\degr, the M-star binary orbit
is seen at a 31\degr angle \citep{bard1987}. 
Unfortunately, the outer 20-year period
orbit has not been resolved visually nor astrometrically yet, 
although in the Hipparcos catalogue it is suspected to be 
an astrometric double (the ``S'' flag in the H61 field).

Our new observations (February to April 2006) were obtained 
almost exactly one whole 20-year period of the outer orbit 
after the
spectroscopy of \citet{bard1987}. Hence we cannot provide
any new insight into the outer orbit. The systemic 
(center of mass) velocity of the contact pair is expected 
to have changed due to mutual revolution by $\pm 6.23$ km~s$^{-1}$
according to the most recent light-time effect (hereafter LITE) 
solution \citep{prib2006}. An evenly distributed spectroscopic 
coverage of the 20-year orbit would enable to unambiguously determine 
masses of all four components in the system, the same way as it was 
done for VW~LMi \citep{ddo11}. 

In the non-Keplerian solution (with proximity effects included) 
of \citet{bard1987}, $K_1 = 124.1 \pm 2.8$ km~s$^{-1}$ is 
significantly smaller than our value of 144.65 km~s$^{-1}$.
It is interesting to note that \citet{hriv1984} 
obtained an even smaller $K_1$ than
\citet{bard1987}, only $108 \pm 2$ km~s$^{-1}$. 
This systematic effect is probably due to the previous use of
the cross-correlation technique and thus inadequate resolution
leading to a stronger influence of the 
third component which is always close to the center-of mass 
velocity of the close binary.
As a result, our mass ratio for the contact pair 
is relatively large compared with the previous results, 
$q = 0.729(7)$.

At 5184~\AA, the light contribution of the fourth component 
(the secondary of the M-dwarf pair)
is only about 1--2\% so that this component is not seen 
in our spectra; therefore, we could determine only a 
single-line orbit for the second pair (Table~\ref{tab4} and
Fig.~\ref{fig5}). The parameters,
$V_0 = -39.67 \pm 0.27$ km~s$^{-1}$, $K_3 = 46.44 \pm 0.38$ 
km~s$^{-1}$, are practically identical with those for the
solution of \citet{bard1987}. The light contribution of the 
third component estimated from the 
triple-Gaussian fits to our BF's around the quadratures of the 
contact binary is about $L_3/(L_1+L_2) = 0.13$, 
which is much higher than that found in the 
photometric analysis of \citet{yakut03}. We note that
because of the different spectral continuum normalization 
levels for slowly and rapidly rotating 
components in spectroscopically multiple
systems, the spectroscopic estimates tend 
to overestimate this ratio.

\subsection{AM~Leo}
\label{amleo}

The contact binary AM~Leo is the brighter component of the visual 
double star ADS~8024 (WDS~J11022+0954)
with the separation of 11.5 arcsec. 
The position angle of the fainter 
companion is 270\degr, i.e.\ along our spectrograph slit, so the
light of both components entered the spectrograph and 
was recorded simultaneously. 
The radial velocity of the visual companion,
$V_3 = -11.08 \pm 0.97$ km~s$^{-1}$, was found to be stable and 
close to the systemic velocity of the eclipsing binary. Due to 
the relatively high brightness of the eclipsing pair, it was a
subject of numerous photometric investigations 
(for references see \citet{hiller2004,alba2005}). 
\citet{hiller2004} analyzed the light curves of the system 
and found $q=0.398$ and the inclination angle 
$i = 86\degr$. While AM~Leo was included in
the Hipparcos mission, the presence of the visual companion 
significantly deteriorated its trigonometric parallax 
determination leading to its large
uncertainty, $\pi = 13.03 \pm 3.64$ mass.

The only spectroscopic orbit was presented by \citet{hriv1993} 
who performed a preliminary solution neglecting proximity 
effects and finding the 
mass ratio of $q = 0.45$ with $M_1 + M_2 = 2.00 \, M_\odot$.
Our solution with $q = 0.459(4)$ is consistent with the above 
determination. The system is seen almost edge-on, so that 
the minimum mass derived by us, 
$M_1 + M_2 = 1.882 \pm 0.018 \, M_\odot$, is close to the
true total mass of the system.

The recent period study of AM~Leo \citet{alba2005} shows possible cyclic
period variations interpreted by the authors as a result of a LITE
caused by an invisible third component with the
minimum mass of $0.18 M_\odot$. The authors estimated 
that this hypothetical body would be about 7 magnitudes
(i.e., more than 600 times) fainter than
the contact binary. This component, if it really exists, 
cannot be identified with the known visual companion on the wide
astrometric orbit. As expected, with the 7 mag.\ difference,
we do not see any persistent feature close to the systemic velocity  
which could be interpreted as being caused by a faint nearby companion.
In fact, in the DDO averaged spectra \citep{dang2006},  
the brightness difference detection limit is about 5.2 magnitudes.

\subsection{V1010~Oph}
\label{v1010oph}

V1010~Oph is a bright ($V_{max} = 6.20$) early spectral type 
(A3V) short-period, almost-contact, 
semi-detached eclipsing binary. Its variability was discovered 
by \citet{stroh1964}.
The published spectroscopic studies of the system 
\citep{guin1977,marg1981,wor1988} found V1010~Oph to be a SB1 system 
with $K_1 \simeq 100$ km~s$^{-1}$. \citet{marg1981} observations
led to a significant orbital eccentricity of $0.23 \pm 0.03$ 
which is highly unexpected for such a close binary and 
indicated a problem in the analysis.
Later observations of \citet{wor1988} gave a small
eccentricity of $e=0.02 \pm 0.02$ which is consistent with
zero due to the biased character of the eccentricity estimates 
(it cannot be negative). 
While the previous velocity semi-amplitudes were consistent, 
the center of mass velocity was discordant, with
$V_0=-15 \pm 3$ km~s$^{-1}$  determined by \citet{marg1981} 
and $V_0=-41 \pm 1.5$ km~s$^{-1}$ determined by \citet{guin1977}. 
The photometric analysis of \citet{leun1977}, based on the assumption
of the Roche model, showed that (i)~the primary eclipse is a transit, 
(ii)~the eclipses are total and (iii)~the system is in marginal contact.

Our BF's of V1010~Oph (Fig.~\ref{fig4}) clearly show the 
secondary component orbiting with a 
large semi-amplitude (see Table \ref{tab2}).  
The well determined mass ratio of $q_{sp} = 0.465 \pm 0.003$
is in agreement with the photometric determination 
$q_{ph} = 0.4891 \pm 0.0016$ of \citet{leun1977}. Note, that this
statement is a qualified one because 
very frequently we see large
discordances between $q_{sp}$ and $q_{ph}$ and 
even in this case the error of $q_{ph}$  must have been strongly 
under-estimated. As expected, we
do not see any indications of a non-zero eccentricity.
It is interesting to note that all radial-velocity
determinations from 2005 (plotted as open symbols in Fig.~\ref{fig2}) 
give a much smaller systemic velocity for the system, by
$\Delta V_0 \simeq -30$ km~s$^{-1}$, possibly indicating that 
the eclipsing pair orbits around a common center of gravity with a 
third star in the system.
The system was included into the Hipparcos mission with the 
resulting parallax of $13.47 \pm 0.83$ mas, so the system has
a very well determined distance. 

\subsection{V2612~Oph}

The variability of V2612~Oph (= NSV~10892 in the General Catalog
of Variable Stars) was first
 suspected by \citet{hilt58}. The author gave 
$V = 9.50$, $B-V = 0.60$ and $U-B = 0.07$. 
$V$-band photometry of \citet{kopp02} showed that it 
is a  contact binary. The authors determined a preliminary 
ephemeris for the primary minimum: 
$Min.I = HJD 2,452,454.7107 + 0.375296\times E$. 
Their light curve is asymmetric with the maximum 
following the primary minimum brighter by about 0.03 mag. 
New minima of V2612~Oph observed by \citet{tas04}, 
gave a large $(O-C)$ shift indicating a
slightly longer orbital period. Therefore in our 
spectroscopic solution, we optimized both 
$T_0$ and $P$ leading to $P$ = 0.375307(3). 

\citet{yang05} analyzed the light curve of 
\citet{kopp02} and found the following parameters 
$q  = 0.323 \pm 0.002$, $i = 65.7 \pm 0.3$ and 
$f = 0.23 \pm 0.04$. 
Our mass ratio $q = 0.286$ is not consistent
with the photometric estimate, as frequently
observed for partially eclipsing system with over-interpreted
light curve analyses. 
$T_0$ in the ephemeris of \citet{kopp02} coincides
with upper conjunction of the less massive component 
so the system is clearly of the W subtype. 
The projected total mass of the system, 
$(M_1 + M_2) \sin^3 i = 1.279 \pm 0.011 \, M_\odot$, 
is consistent with our spectral type estimate, F7V,
for a moderately low value of the orbital inclination.

V2612~Oph is located in the outskirts of the intermediate 
age galactic cluster NGC~6633. The star was included 
in the four color photometry of NGC~6633 by \citet{schmidt76}
who found $(b-y) = 0.382$ and determined a large value for the
interstellar reddening of $E(b-y) = 0.472$, which is inconsistent
with the average cluster reddening of $E(b-y) = 0.124 \pm 0.017$.
Similarly to \citet{hilt58}, the author did not accept 
the membership of the star in NGC~6633.

V2612~Oph was not observed by Hipparcos satellite, 
and its trigonometric parallax is unknown so the cluster 
membership cannot be reliably verified. Using the \citet{rd1997} 
absolute magnitude calibration assuming F7V spectral type, 
we obtain $M_V = 3.52$. With the distance 
modulus of NGC~6633 of  $V - M_V = 7.71$ mag \citep{kharch05} 
the system should be as faint as $V_{max} = 11.26$. 
Hence, it seems that V2612~Oph is in front of the cluster,
although this is entirely inconsistent with the supposedly
very large $E(b-y)$ reddening value.

The proper motion of V2612~Oph, 
$\mu_{\alpha} \cos \delta = 57.2 \pm 2.1$ mas/year and 
$\mu_{\delta} = 23.3 \pm 2.1$ mas/year \citep{Tycho2}, 
does not correspond to the mean NGC~6633
motion of $\mu_{\alpha} \cos \delta = 0.10$ mas/year 
and $\mu_{\delta} = -2.0$ mas/year \citep{kharch05}. 
On the other hand, the center of mass velocity of V2612~Oph, 
$V_0 = -25.59 \pm 0.44$ km~s$^{-1}$ is close to 
mean radial velocity of the cluster of  
$V_R = -25.43$ km~s$^{-1}$, as given by \cite{kharch05}. 
Recently, high precision photometry of the cluster 
performed by \citet{hid2005} has led to
detection of several variable stars in NGC~6633. In 
particular, a W~UMa variable V7 with a similar 
period of $P = 0.38673$ days, at $V_{max} = 12.8$,
is over 3 magnitudes fainter than V2612~Oph. Hence
we reject the membership of V2612~Oph to NGC~6633.

\subsection{XX~Sex}
\label{xxsex}

The photometric variability of XX~Sex (HD~89027) 
was found on the Stardial images \citep{wils2003}. 
The authors determined an
approximate ephemeris for the primary minima $Min.I =
HJD 2452314.79 + 0.54011\times E$. The ASAS-3 light curve shows 
that XX~Sex is a totally eclipsing system with rather 
different depths of the minima. The orbital period in ASAS is 
slightly improved to 0.540111 days. 
No high-precision photometry of XX~Sex has been published yet.

Our independent spectral type estimate of F3V is slightly 
later than F0, as given in Simbad Astronomical database. 
XX~Sex was not included in the Hipparcos mission, 
hence its trigonometric parallax is unknown. 
During our spectroscopic observations 
we noted a faint visual companion to XX~Sex separated by about 3 arcsec
in the NW direction.

The upper spectroscopic conjunction of the more massive component 
observed by us coincides in phase with the deeper minimum in the ASAS-3 
photometry indicating that the system is of the A-type. 
This is further supported by the low mass ratio  determined by
us, $q = 0.100 \pm 0.002$ and by the relatively long orbital period. 
For such a small mass ratio, eclipses remain total for a wide range of
inclinations down to as low as $70\degr$. This is actually indicated by 
the relatively small projected total mass of 
$(M_1+M_2) \sin^3 i = 1.153 \pm 0.026 \, M_\odot$ 
for the spectral type of F3V.

In the analysis of the broadening functions we noted
that while the peak of the secondary component is usually well 
defined around the second quadrature (phase 0.75), 
where it was sufficiently separated from the primary-component peak, 
its profile is flat and poorly defined around the first quadrature. 
These unexplained shape variations in the secondary
component signature resulted in rejection of four RV 
determinations for this component.

\subsection{W~UMa}
\label{wuma}

The prototype contact binary W~UMa has been intensively observed 
since its discovery in 1903 \citep{mull1903}. 
The contact binary is the brighter ($\Delta m = 4.4$) component of the 
visual pair ADS~7494 (WDS J09438+5557) with the separation of 6.4 arcsec.
The physical association of the components has not been
yet demonstrated. The Hipparcos Catalogue lists W~UMa as a suspected 
astrometric binary with ``S'' flag in the H61 field.

Apart from numerous photometric observations and studies 
\citep{linn1991}, the system was observed 
several times spectroscopically \citep{mclean1981,wuma}. 
The spectroscopic elements of the system are still poorly known with 
large differences in the center of mass velocity ranging from 
$V_0=0$ km~s$^{-1}$ \citep{binn1967}
to $V_0=-43$ km~s$^{-1}$ \citep{popp1950}; it is unclear
if these differences can be explained by very different
and slowly improving methods of radial velocity determinations
for contact binaries. 
In addition, W~UMa shows unexplained, irregular orbital period changes 
which probably indicate simultaneous action of several 
mass and angular-momentum transfer processes within the contact
system.

Our new spectroscopic orbit is based on 36 high-precision RV
measurements extracted from very-well 
defined BF's (see Fig.~\ref{fig4}). 
The resulting parameters (Table~\ref{tab2})
clearly supersede all previous determinations. 
The spectroscopic elements are well 
within previous determinations uncertainties, as given 
in the last two studies \citep{mclean1981,wuma}.
During our observations covering 16 days we didn't observe 
any center-of-mass velocity changes. 

The trigonometric parallax of the system, $\pi = 20.17 \pm 1.05$, 
is very well defined. The absolute visual magnitude determined 
from the period-color-luminosity 
relation of \citet{rd1997}, $M_V = 3.86$, is rather severely
inconsistent with the magnitude found from the parallax, 
$M_V = 4.82 \pm 0.11$. To obtain a reasonable accord, a substantial 
amount of reddening ($A_V = 0.96$) or a much later spectral type --
perhaps as late as K1V -- would be required; 
both explanations are equally unlikely.
This major discrepancy is entirely unexplained and puzzling
for such a well observed contact binary.

\subsection{XY~UMa}
\label{xyuma}

XY UMa is a highly chromospherically active system
with an exceptionally short orbital period for a detached binary 
of only 0.479 days. The photometric variability of 
XY~UMa was first noted by \citet{geyer1955}. 
The binary has been the subject to extensive photometric studies;
for references see \citet{prib2001}. The spectroscopic orbit 
of the primary  component was first obtained by the CCF technique 
by \citet{rain1991}. To detect and measure the RV's
for the secondary, \citet{pojm1998} applied a sophisticated modeling 
of the near-infrared spectra in the region of the Ca~II 
infra-red triplet. This led to the 
following parameters $V_0 = -10.5 \pm 1.0$ km~s$^{-1}$, 
$K_1 = 122.5 \pm 1.0$ km~s$^{-1}$, $K_2 = 202 \pm 6$ km~s$^{-1}$.
The chromospheric emission filling the $H_\beta$ and $H_\alpha$ 
lines was later studied by \citet{ozer2001}. 

The system was suspected 
to be a member of a multiple system \citep{prib2001} 
with the third-body orbital period of about 30 years.
The H61 field of the Hipparcos Catalog contains flag ``S'' 
(as described in the catalog, ``suspected non-single''),
i.e., a plausible astrometric orbital solution for XY~UMa
was found. While our BF's do not show any trace of the third component, 
we noted during the current observations 
a faint ($\Delta m \approx 3$ mag) visual companion 
in the NW direction about 2--3 
arcsec from XY~UMa.  A spectrum of the companion taken on 
March 24, 2006 indicates that it may be a binary system,
although contamination by the light of XY~UMa is not excluded.

Our new spectroscopic observations consist of two runs: 
In the spring of 2005, we observed spectra centered at 
6290~\AA, in a spectral window which included a telluric 
molecular feature (later removed in the BF extraction), while
in the spring of 2006 we used the standard setup around Mg~I 
5184~\AA\ triplet. Neither of the runs revealed any obvious 
signatures of the secondary component so, at first, 
we treated the orbit as that of a single-lined binary (SB1). 
An orbit based only on the 2006
observations, $V_0 = -7.68 \pm 0.24$ km~s$^{-1}$ 
and $K_1 = 124.74 \pm 0.28$ km~s$^{-1}$ 
utilized 61 spectra (excluding spectra
obtained within $\pm 0.09$ in phase around the primary 
eclipse). When we arranged the spectra in phase and smoothed them
in the phase domain, a faint feature of the secondary component
became visible (Fig.~\ref{fig6}). Its semi-amplitude was about 
$K_2 \simeq 178$ km~s$^{-1}$, which is
markedly smaller than 202 km~s$^{-1}$,  
found by \citet{pojm1998}. The discrepancy may 
be caused by the dominance of the reflection effect at 
5184~\AA\  so that the effective line center on the secondary is shifted 
towards its irradiated hemisphere.  
We note that our BF's are very well defined for the primary component.
They clearly show a dark, relatively small and well 
localized spot visible 
around the second quadrature which migrates through the stellar
profile following the stellar rotation of the primary component. 
A weaker similar spot was recorded around the first 
quadrature in some of the spectra.

The Hipparcos parallax of the system, $15.09 \pm 1.48$ mas, and the 
maximum visual magnitude $V = 9.62$ give the absolute visual 
magnitude $M_V = 5.51 \pm 0.22$ corresponding to a single 
G8V main-sequence star. The light contribution of the secondary 
companion is just a few percent in the $V$ passband.

The spectral type of XY~UMa was estimated by analyzing average
spectrum of the system. The best template to fit the average spectrum
was found to be K0V
(the next available templates were G8V and K4V). This template 
was the best even for individual spectra observed during the eclipses.
The published spectral types of the components are 
G3V + K4-5V \citep{prib2001}, which appear to be much too early.

\section{SUMMARY}
\label{summary}

With the new ten short-period binaries, this paper brings the
number of the systems studied at the David Dunlap Observatory to
one hundred and ten. The systems presented in this paper include 
(1)~the quadruple system XY~Leo consisting of a contact binary and
a BY~Dra-type close binary consisting of 
two M-type dwarfs, (2)~the very
close, detached, chromospherically active system XY~UMa,
(3)~V1010~Oph which is probably a detached 
or semi-detached SB2 system. 
The remaining seven SB2 binaries are all contact
ones: OO~Aql, CC~Com, V345~Gem, AM~Leo, V2612~Oph, XX~Sex and W~UMa. 
Six systems of this group were observed spectroscopically before:
OO~Aql, CC~Com, XY~Leo, AM~Leo, W~UMa, and XY~UMa, but our new
data are of higher quality than in any of the previous studies. 

Companions to the close binaries appear to be present in
V345~Gem, XY~Leo, AM~Leo, XX~Sex and XY~UMa, but all have been
recognized as such before except for XX~Sex and XY~UMa where the
faint companions are new detections.
The case for the physical association of the visual companion 
to W~UMa still remains open. 

We point out that the red color of OO~Aql is unexplained,
unless it is very heavily reddened.
We also note a large discrepancy in the absolute magnitude of
W~UMa between the predicted by the
simple period--color--luminosity calibration
and the one derived from the parallax. 

\acknowledgements

We express our thanks to  Matt Rock, Tomasz Kwiatkowski for the
observations and to
Wojtek Pych for providing his cosmic-ray removal code.
Support from the Natural Sciences and Engineering Council of Canada
to SMR and SWM and from the Polish Science Committee (KBN
grants PO3D~006~22 and P03D~003~24) to WO 
is acknowledged with gratitude. The travel of TP to
Canada has been supported by a Slovak Academy of Sciences 
VEGA grant 2/7010/7.
TP appreciates the hospitality and support of the local staff
during his stay at DDO. 
The research made use of the SIMBAD database, operated at the CDS,
Strasbourg, France and accessible through the Canadian
Astronomy Data Centre, which is operated by the Herzberg Institute of
Astrophysics, National Research Council of Canada.
This research made also use of the Washington Double Star (WDS)
Catalog maintained at the U.S. Naval Observatory.

\clearpage

\noindent
Captions to figures:

\bigskip

\figcaption[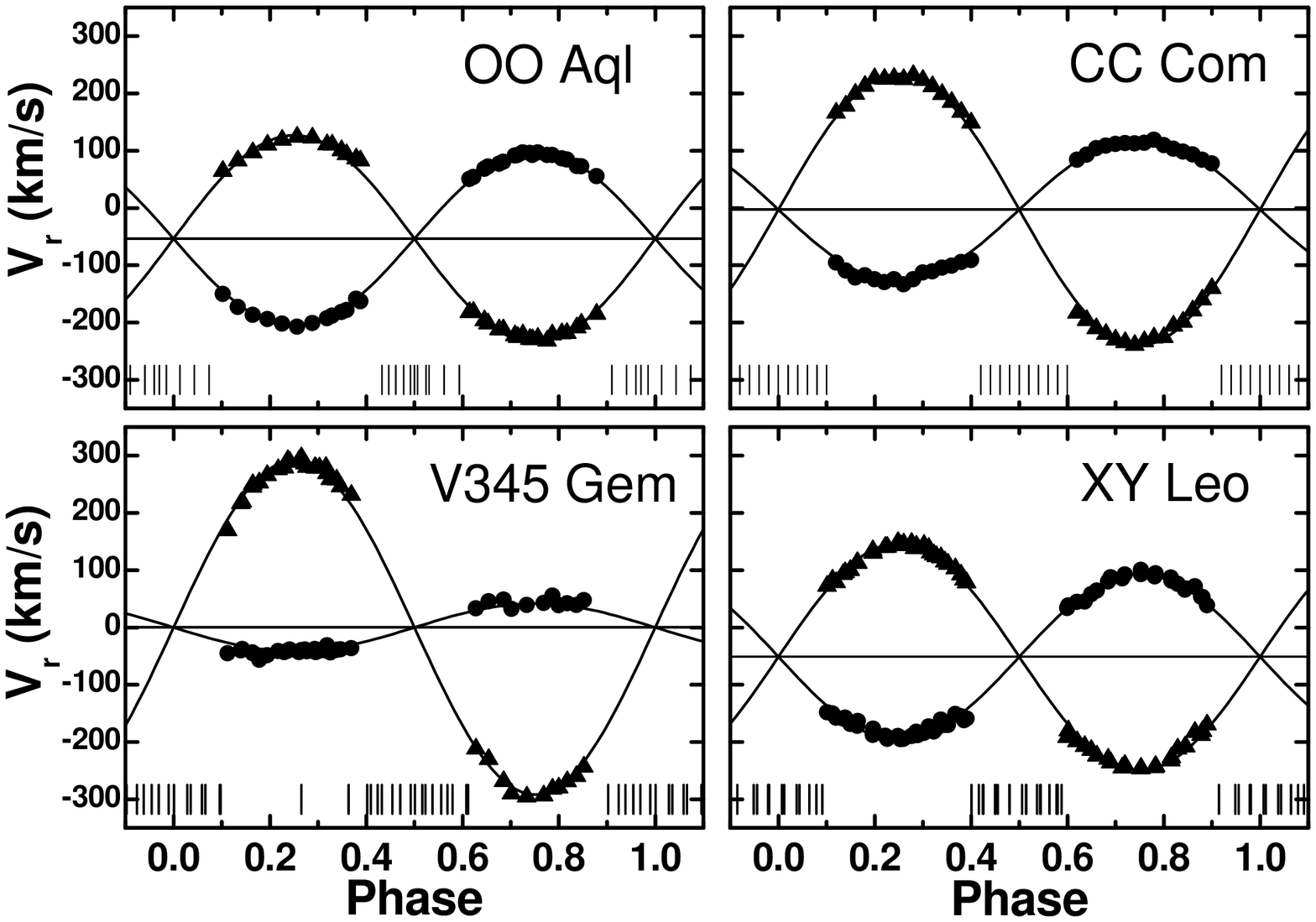] {\label{fig1} Radial velocities of the
systems OO~Aql, CC~Com, V345~Gem and XY~Leo are plotted in
individual panels versus the orbital phases. The lines give the
respective circular-orbit (sine-curve) fits to the RV's.
While all four systems are contact binaries, V345~Gem and XY~Leo are 
members of multiple systems. The circles and triangles in this and 
the next two figures correspond to components with velocities 
$V_1$ and $V_2$, as listed in Table~\ref{tab1}, respectively. The 
component eclipsed at the minimum corresponding to $T_0$ (as given 
in Table~\ref{tab2}) is the one which shows negative velocities for 
the phase interval $0.0 - 0.5$ and which is the more massive one. 
Short marks in the lower parts of the panels show phases of available 
observations which were not used in the solutions because of the 
excessive spectral line blending. 
}

\figcaption[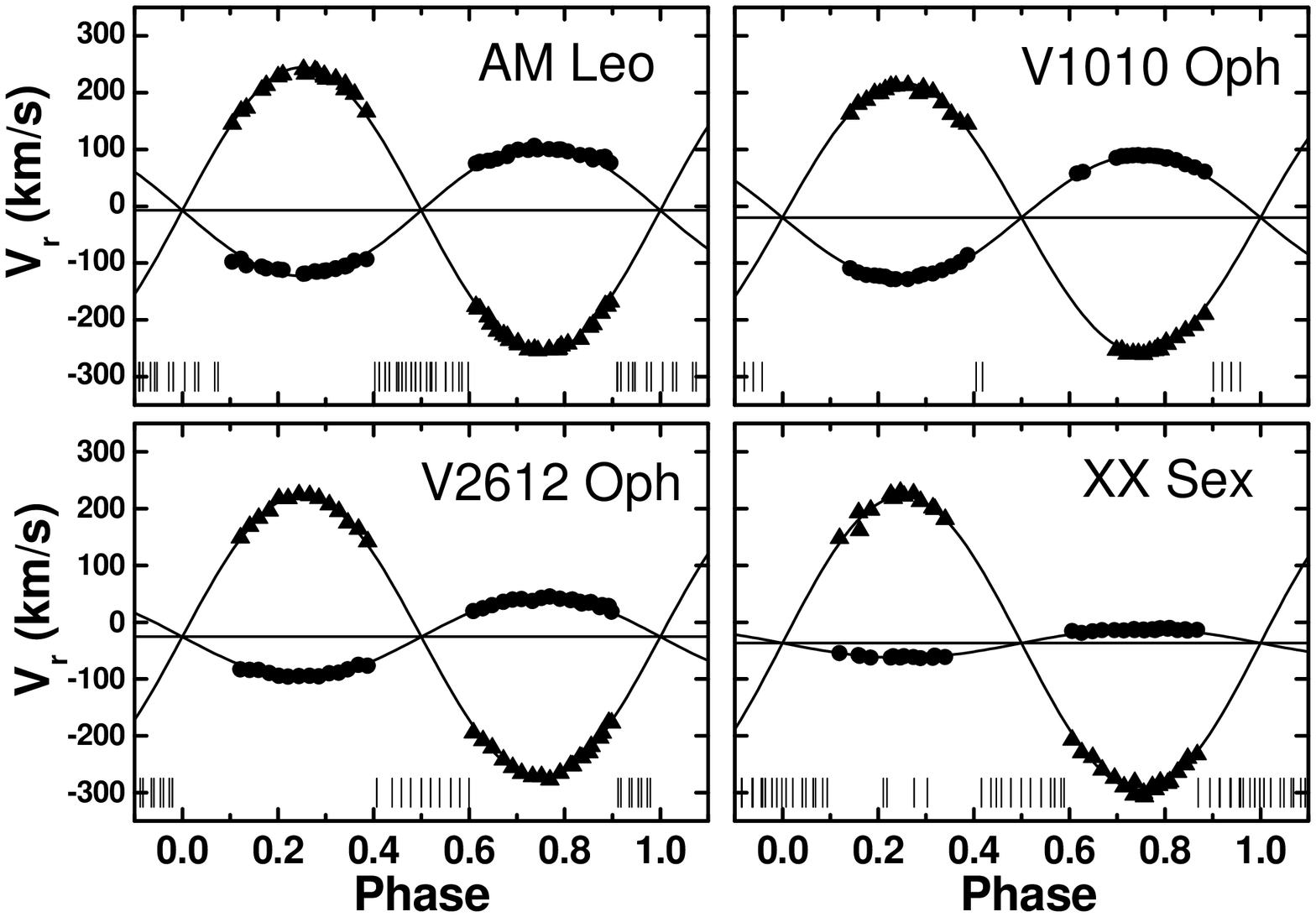] {\label{fig2} The same as for
Figure~\ref{fig1}, but for AM~Leo, V1010~Oph, V2612~Oph, and XX~Sex.
While V1010~Oph is double-lined detached or
semi-detached binary, AM~Leo, V2612~Oph, and 
XX~Sex are contact binaries. Open symbols correspond to the observations 
not used for the spectroscopic orbit determination.
}

\figcaption[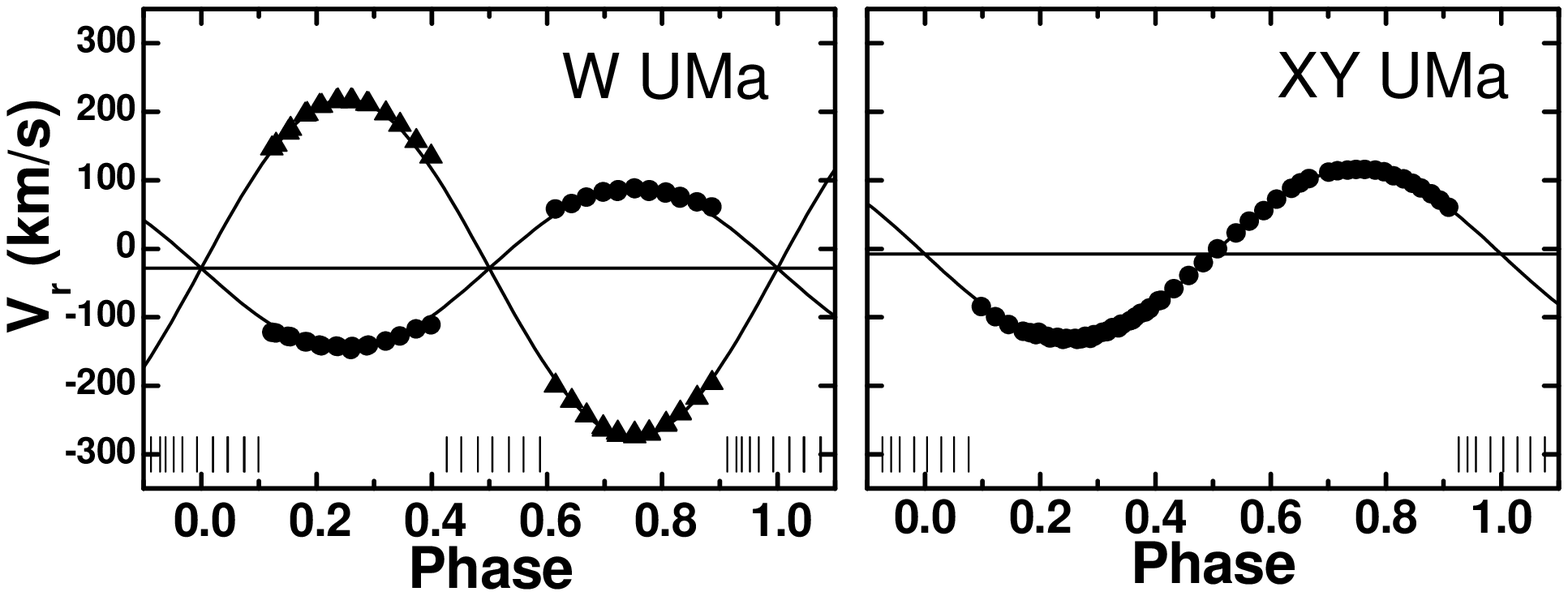] {\label{fig3} The same as for
Figures~\ref{fig1} and \ref{fig2}, but for the two remaining systems
W~UMa, and XY~UMa. While W~UMa is the prototype contact binary,
XY~UMa is a very close, but detached binary.}

\figcaption[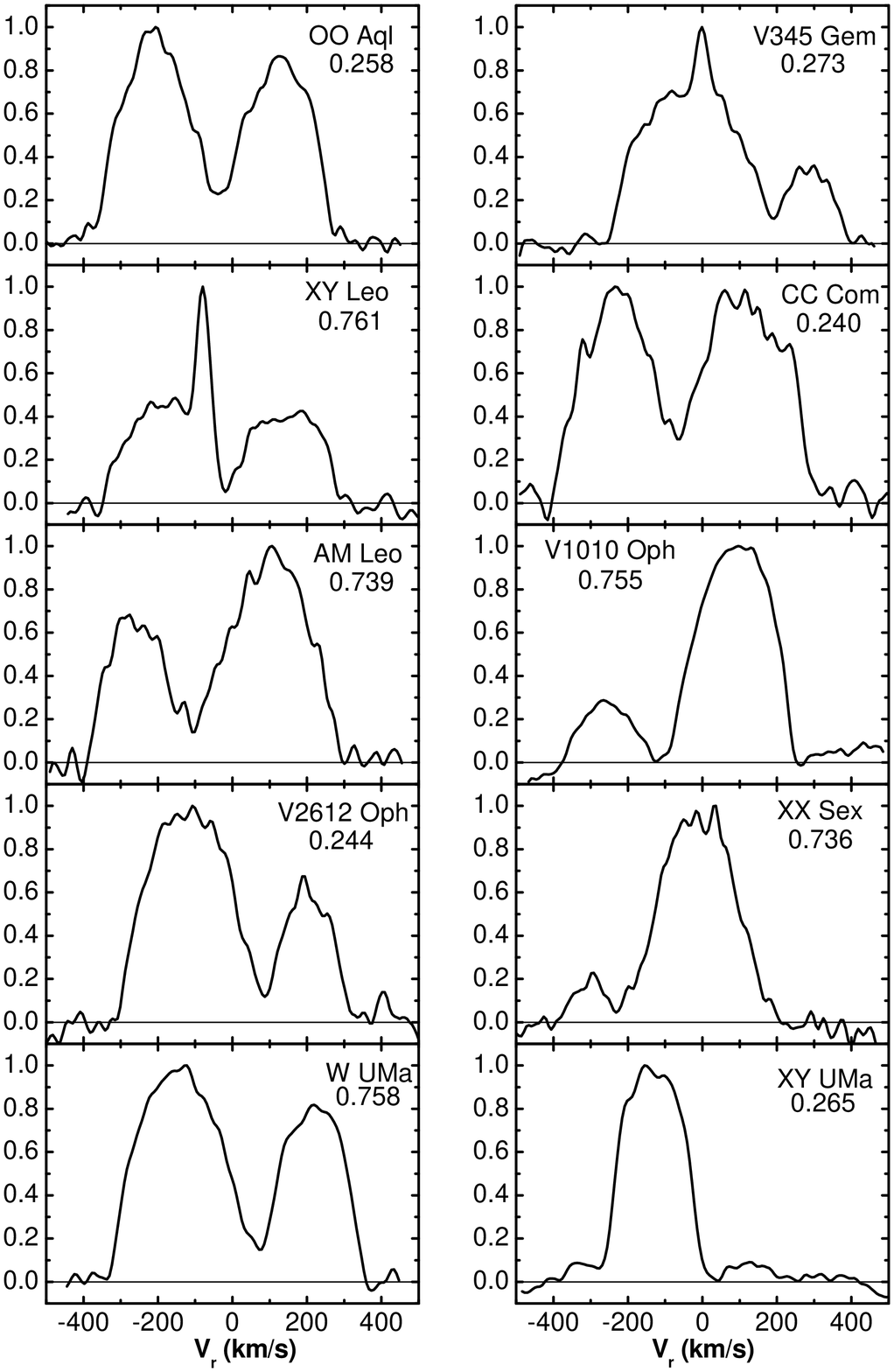] {\label{fig4}
The broadening functions (BF's) for all ten systems of this
group, selected for phases close to 0.25 or 0.75.
The phases are given by numbers in individual panels.
XY~Leo is quadruple system composed of the contact eclipsing binary
and of the detached, non-eclipsing, close binary with the orbital period
$P \approx 0.805$ days and with only one component visible as a
relatively sharp peak in the BF (its orbit is shown in Fig~\ref{fig5}).
The third star feature in the BF of the contact
binary V345~Gem is also clearly visible. All panels have the
same horizontal range, $-500$ to $+500$ km~s$^{-1}$.}

\figcaption[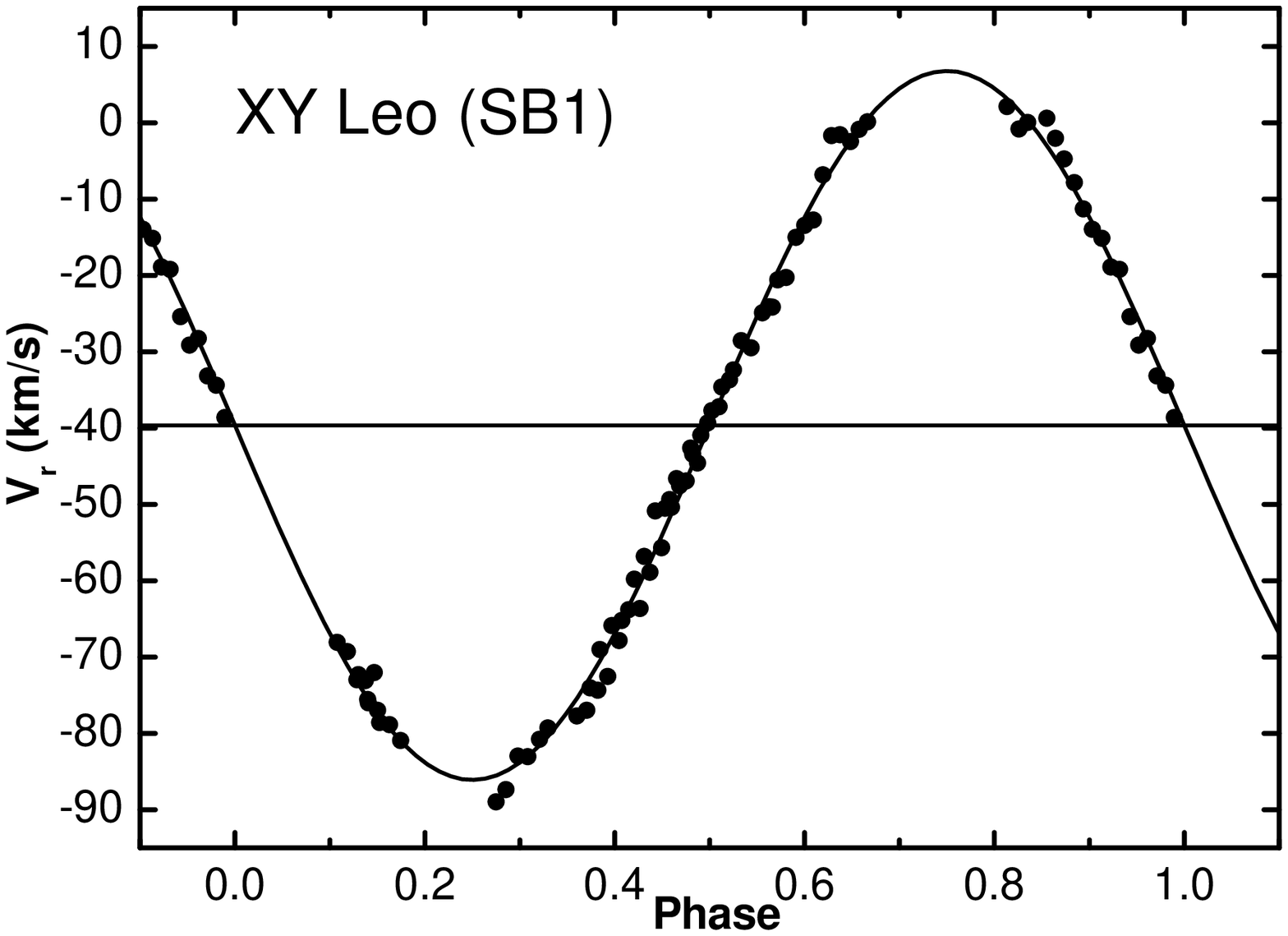] {\label{fig5}
The radial velocities of the third component of XY~Leo and
the corresponding fit to its orbital motion with the period
of 0.805 days.}

\figcaption[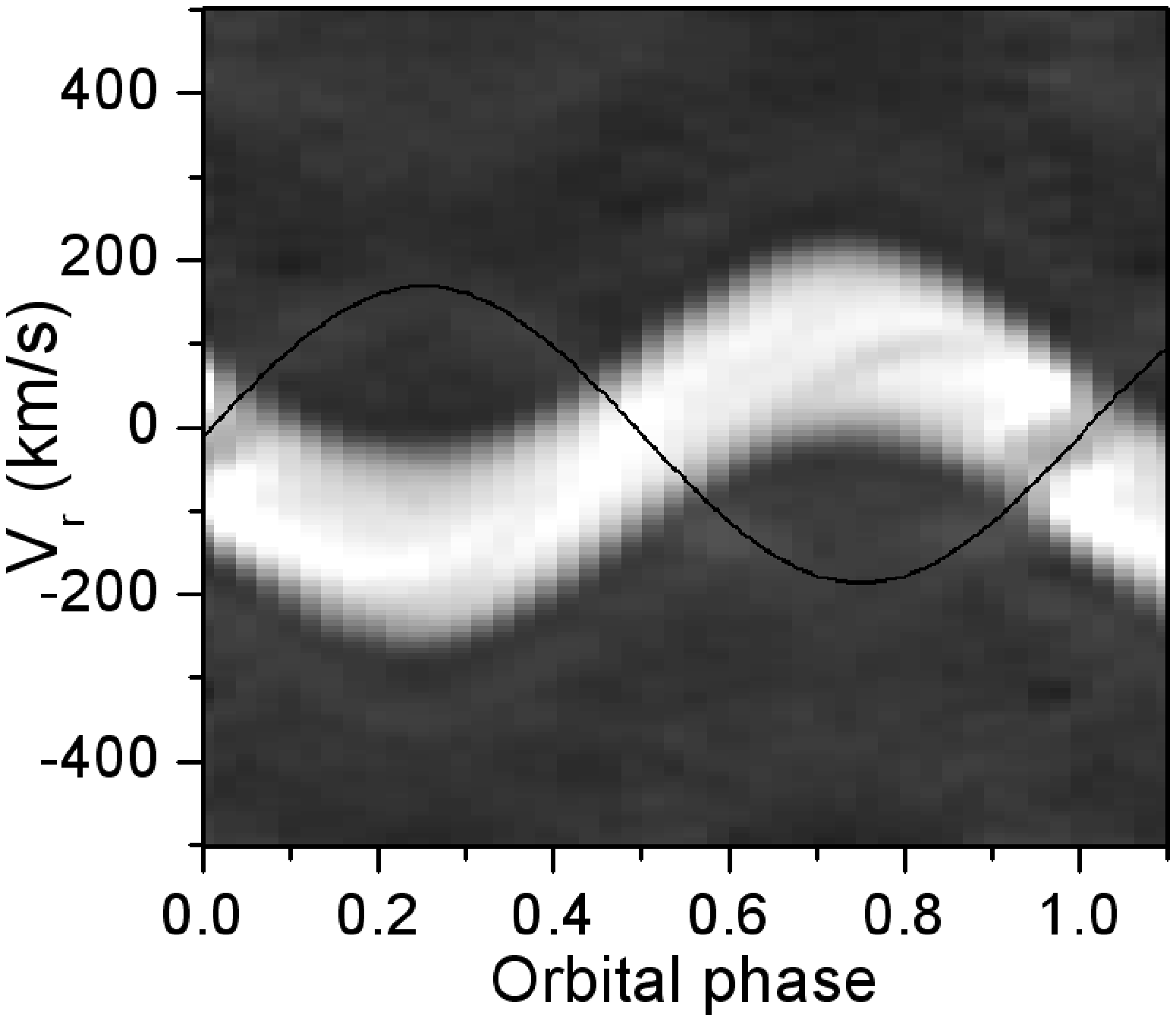] {\label{fig6}
A gray-scale plot of the phase-domain averaged broadening
functions of XY~UMa showing faint features of the secondary
companion, best visible after the secondary minimum. During the
primary eclipse the secondary component is clearly visible as a moving
dark feature within the primary component profile. Note also
the dark spots on the primary best visible in phases 0.7 – 0.8.
An orbit of the secondary component, 
as observed in our spectral window at 5184 \AA,
is plotted by a solid line.}

\clearpage
\plotone{f1.eps}
% fig.1

\clearpage
\plotone{f2.eps}
% fig.2

\clearpage
\plotone{f3.eps}
% fig.3

\clearpage
\epsscale{.8}
\plotone{f4.eps}
\epsscale{1}
% fig.4

\clearpage
\plotone{f5.eps}
% fig.5

\clearpage
\plotone{f6.eps}
% fig.6

%----------------------------------------------------------------------
\begin{deluxetable}{lrrrrr}

%\small
\tabletypesize{\footnotesize}
%\tabletypesize{\scriptsize}

\tablewidth{0pt}
%\tablewidth{320pt}
\tablenum{1}

\tablecaption{DDO radial velocity observations (the full
table is available only in electronic form
\label{tab1}}

\tablehead{
\colhead{HJD--2,400,000}  &
\colhead{~$V_1$}          &
\colhead{~~$W_1$}         &
\colhead{~$V_2$}          &
\colhead{~~$W_2$}         &
\colhead{Phase}          \\
                          &
\colhead{[km s$^{-1}$]}   &
                          &
\colhead{[km s$^{-1}$]}   &
                          & \\
}
\startdata
%\sidehead{\bf OO~Aql} \\
 53590.6265  &   0.00~ &  0.00 & $   0.00$~ &  0.00 &  0.4984 \\
 53590.6421  &   0.00~ &  0.00 & $   0.00$~ &  0.00 &  0.5292 \\
 53590.6578  &   0.00~ &  0.00 & $   0.00$~ &  0.00 &  0.5602 \\
 53590.6738  &   0.00~ &  0.00 & $   0.00$~ &  0.00 &  0.5917 \\
 53590.6893  &  53.83~ &  2.65 & $-181.89$~ &  1.05 &  0.6223 \\
 53590.7049  &  72.29~ &  5.60 & $-201.81$~ &  0.73 &  0.6531 \\
 53590.7208  &  80.82~ &  0.54 & $-212.17$~ &  0.45 &  0.6845 \\
 53590.7362  &  92.63~ &  0.74 & $-225.73$~ &  0.46 &  0.7149 \\
 53590.7516  &  92.74~ &  0.69 & $-229.77$~ &  0.34 &  0.7453 \\
 53590.7669  &  92.44~ &  0.65 & $-232.10$~ &  0.30 &  0.7754 \\
\enddata

\tablecomments{The table gives the RV's $V_i$ and
associated weights $W_i$ for observations described in the paper.
The first 10 rows of the table for the first program star, OO~Aql,
are shown. Observations leading to entirely inseparable broadening
function peaks are given zero weight; these observations may be
eventually used in more extensive modeling of broadening
functions. The RV's designated as $V_1$ correspond to
the more massive component; it was always the component eclipsed
during the minimum at the epoch $T_0$ (this not always corresponds
to the deeper minimum and photometric phase 0.0). The phases
correspond to $T_0$ and periods given in Table~2.}

\end{deluxetable}

%-------------------------------------------------------------------------
\begin{deluxetable}{lccrrrccc}

%\tabletypesize{\small}
\tabletypesize{\scriptsize}

\pagestyle{empty}
\tablecolumns{9}

%\tablewidth{660pt}
\tablewidth{0pt}

\tablenum{2}
\tablecaption{Spectroscopic orbital elements \label{tab2}}
\tablehead{
   \colhead{Name} &                % 1
   \colhead{Type} &                % 2
   \colhead{Other names} &         % 3
   \colhead{$V_0$~~~} &            % 4
   \colhead{$K_1$~~~} &            % 5
   \colhead{$\epsilon_1$~} &       % 6
   \colhead{T$_0$ -- 2,400,000} &  % 7
   \colhead{P (days)} &            % 8
   \colhead{$q$}          \\       % 9
   \colhead{}     &                % 1
   \colhead{Sp.~type}    &         % 2
   \colhead{}      &               % 3
   \colhead{} &                    % 4
   \colhead{$K_2$~~~} &            % 5
   \colhead{$\epsilon_2$~} &       % 6
   \colhead{$(O-C)$(d)~[E]} &      % 7
   \colhead{$(M_1+M_2) \sin^3i$} & % 8
   \colhead{}                      % 9
}
% format template
% (1) name, sp & (2) type           & (3) name HD/BD   & (4) V0   & (5) K1 &
% (6) errV1    & (7) T0-2,400,000   & (8) P            & (9) q    \\
% (1)          & (2) sp type        & (3) BD/Hip       & (4)      & (5) K2 &
% (6) errV2    & (7) O-C ~ [E]      & (8) (M1+M2)sin3i & (9)      \\
\startdata

%OO Aql, A-type checked, done
OO~Aql     & EW(A)             & HD~187183   & $-53.71$(0.61)  & 153.03(0.93) &
      4.08 & 53,606.0845(6)    & 0.5067932   & 0.846(7)        \\ %32 spectra from 2005/2006
           & F9V               & BD+08~4224  &                 & 180.81(1.14) &
      7.35 & $+0.0008$~[+2,182]& 1.954(19)   &                 \\[1mm]

%CC~Com, W-type checked
CC~Com     & EW(W)             & GSC~1986-2106 & $-2.89$(0.74)   & 124.83(1.34) &
     5.72  & 53,822.2339(3)    & 0.2206860   & 0.527(6)        \\ % 28/27 averaged BFs from 2006
           & K4/5              &             &                 & 237.00(1.09) &
     5.54  & $-0.0004$[+5,990.5]& 1.083(12)  &                 \\[1mm]

%V345 Gem, W-type checked
V345~Gem   & EW(W)             & HD~60987    & $+0.03$(0.68)   &  41.54(0.96) &
     5.31  & 53,802.8329(3)    & 0.2747690   & 0.142(3)        \\ % 47 spectra from 2006
           & F7V               & HIP~37197   &                 & 291.75(1.26) &
     6.90  & $+0.0505$[+4,740.5]& 1.054(13)  &                 \\[1mm]

%XY Leo, W-type checked
XY~Leo     & EW(W)             & HIP~49136   & $-51.24$(0.64)  & 144.65(1.10) &
     6.95  & 53,812.1951(3)    & 0.2840978   & 0.729(7)        \\ % 54 spectra from 2006
           & (K0V)             & BD+18~2307  &                 & 198.41(1.11) &
     6.92  & $+0.0022$[+4,618.5]& 1.188(12)  &                 \\[1mm]

%AM Leo, W-type checked 
AM~Leo     & EW(W)             & HIP~53937   & $-7.25$(0.62)   & 115.56(0.97) &
      6.49 & 53,787.5742(12)   & 0.3657989   & 0.459(4)        \\ %45/47 spectra from 2006
           & F5V               & BD+10~2234  &                 & 251.98(1.17) &
           & $+0.0003$[+3,519.5]& 1.882(18)  &                \\[1mm]

%V1010 Oph, only data from 2006 used
V1010~Oph  & EB(SB2)           & HD~151676   & $-19.92$(0.38)  & 110.46(0.45) &
     2.68  & 53,825.7086(19)   & 0.6614168   & 0.465(3)        \\ % 34/31 spectra from 2006
           & A7V               & HIP~82339   &                 & 237.33(1.44) &
     6.48  & $+0.0009$~[+2,004]& 2.883(30)   &                 \\[1mm]

%V2612~Oph, W-type O.K.
V2612~Oph  & EW(W)             & HD~170451   & $-25.59$(0.44) &  71.33(0.66)  &
    3.66   & 53,846.9204(3)    & 0.375307(3) & 0.286(3)       \\ % 34 spectra used
           & F7V               & BD+6 3809   &                & 249.09(0.89)  &
    4.04   & $+0.0492$~[+3,709.5]& 1.279(11) &                \\ [1mm]

%XX Sex, A-type much more probable
XX~Sex   & EW(W)             & HD~89027   & $-36.75$(0.39)  &  25.80(0.45) &
      2.28 & 53,824.4139(9)    & 0.540110    & 0.100(2)       \\ %33 spectra from 2006
           & F3V               & BD-05~3027  &                & 258.51(1.54) &
      7.78 & $+0.0164$~[+2,795]& 1.286(20)   &                \\[1mm]

%W UMa, W-type O.K.
W~UMa      & EW(W)             & HD~83950    & $-28.40$(0.48) &  119.21(0.68) &
     4.90  & 53,804.8472(3)    & 0.33363487  & 0.484(3)       \\ % 36/35 spectra from 2006
           & F5V               & HIP~47727   &                &  246.30(0.87) &
     3.40  & $-0.0006$[+3,910.5]& 1.688(12)  &                \\ [1mm]

%XY UMa
XY~UMa     & EB(SB2:)          & HD~237786   &  $-7.68$(0.24) & 124.74(0.28)  &
     1.67  & 53,821.6344(2)    & 0.4789961   &                \\ % 61 spectra from 2006, 5184 A
           & K0V               & HIP~44998   &                &               &
           & $+0.0000$~[+2,759]&             &                \\[1mm]

\enddata

\tablecomments{The spectral types given in the second column 
relate to the combined spectral type of all components in the 
system; they are given in parentheses if taken from the 
literature, otherwise they are new. 
The convention of naming the binary components in the 
table is that the more massive star is marked by 
the subscript ``1'', so that the  mass ratio is 
defined to be always $q \le 1$. The standard 
errors of the circular  solutions in the table are 
expressed in units of last decimal places quoted; they 
are given in parentheses after each value. 
The center-of-mass velocities ($V_0$), 
the velocity amplitudes ($K_i$) and the standard unit-weight 
errors of the solutions ($\epsilon$) are all expressed in 
km~s$^{-1}$. The spectroscopically determined 
moments of primary or secondary minima are given by $T_0$; 
the corresponding $(O-C)$ deviations (in days) have been 
calculated from the available prediction on 
primary minimum, as given in the text, using the 
assumed periods and the number of 
 epochs given by [E]. 
The values of $(M_1+M_2)\sin^3i$ are in the solar mass units.\\ 
Ephemerides (HJD$_{min}$ -- 2,400,000 + period in days) 
used for the computation of the $(O-C)$ residuals: \\
 OO~Aql:    52500.2610 + 0.5067932;  CC~Com:    52500.2158 + 0.22068583; \\
 V345~Gem:  52500.24   + 0.274769;   XY~Leo:    52500.0872 + 0.2840978; \\
 AM~Leo:    52500.1452 + 0.3657989;  V1010~Oph: 52500.231  + 0.661414; \\
 V2612~Oph: 52454.7107 + 0.375296;   XX~Sex:    52314.79   + 0.54011; \\  
 W~UMa:     52500.1693 + 0.3336347;  XY~UMa:    52500.0844 + 0.4789960.
}
\end{deluxetable}

%----------------------------------------------------------------------
\begin{deluxetable}{lrr}

%\small
\tabletypesize{\footnotesize}
%\tabletypesize{\scriptsize}

\tablewidth{0pt}
%\tablewidth{320pt}
\tablenum{3}
\tablecolumns{4}

\tablecaption{Radial velocity observations (the full
table is available only in electronic form) of visual companion
to V345~Gem and third component in XY~Leo. The radial
velocities of V345~Gem until $HJD~2,453,785$ were derived from
spectra where the single, dominant component was centered on the 
spectrograph slit. After $HJD~2,453,788$ the spectrograph slit
was centered on the fainter eclipsing binary and the radial
velocities were determined by Gaussian profile fitting to the BFs
with the light contamination $L_3/(L_1+L_2)>0.10$. \label{tab3}}
\tablehead{
\colhead{HJD--2,400,000} & \colhead{~V$_3$}        \\
                         & \colhead{[km s$^{-1}$]} \\
}
\startdata
 53780.81180 & $-0.884$ \\
 53780.82239 & $-0.987$ \\
 53780.82978 & $-0.787$ \\
 53781.79724 & $-1.586$ \\
 53781.80200 & $-1.760$ \\
 53781.81979 & $-1.672$ \\
 53781.82691 & $-1.783$ \\
 53785.51372 & $-2.312$ \\
 53785.52096 & $-2.789$ \\
 53785.52816 & $-3.365$ \\
\enddata

\tablecomments{The table gives the RV's for
the visual companion to V345~Gem. The typical 10 rows of the table
are shown}
\end{deluxetable}

%Maybe sigma of the Gaussian fit should be also given...

%----------------------------------------------------------------------
\begin{deluxetable}{lrr}

%\small
\tabletypesize{\footnotesize}
%\tabletypesize{\scriptsize}

\tablewidth{0pt}
%\tablewidth{320pt}
\tablenum{4}

\tablecaption{Spectroscopic orbital elements of the circular orbit of the 
second non-eclipsing SB1 binary in the quadruple system XY~Leo \label{tab4}}

\tablehead{
\colhead{Parameter} &  & \colhead{error} \\
}
\startdata
$P_{34}$ [days]          & 0.80476  &   0.00003    \\
$e_{34}$                 & 0.00     &   --         \\
$\omega$ [rad]           & 1.5708   &   --         \\
$T_0$ [HJD]              & 2,453,814.5286 & 0.0008 \\
$V_0$ [km~s$^{-1}$]      & $-$39.67 &   0.27       \\
$K_3$ [km~s$^{-1}$]      &    46.44 &   0.38       \\
$a_3 \sin i$ [R$_\odot$] &  0.738   &   0.006      \\
$f(m)$ [M$_\odot$]       & 0.0084   &   0.0002     \\
\enddata

\tablecomments{The table gives spectroscopic elements of the second
 binary in XY~Leo: orbital period ($P_{34}$), eccentricity ($e_{34}$),
 longitude of the periastron passage ($\omega$), time of the periastron
 passage ($T_0$), systemic velocity ($V_0$), semi-amplitude of the
 RV changes ($K_3$), semi-major axis of the relative orbit ($a_3 \sin i$),
 mass-function ($f(m)$). The elements were obtained assuming circular
 orbit.}

\end{deluxetable}
\end{document}